\documentclass[11pt,twoside]{article}

\usepackage{asp2021}

\aspSuppressVolSlug
\resetcounters

\begin{document}

\title{Data Formats and Visualisation BoF}

\author{Keith Shortridge}
\affil{K\&V, Sydney, NSW, Australia; \email{KeithShortridge@gmail.com}}

\paperauthor{KeithShortridge}{KeithShortridge@gmail.com}{}{KandV}{}{Sydney}{NSW}{2060}{Australia}



\begin{abstract}
ADASS used to hold a regular FITS BoF (Birds of a Feather meeting). As other data formats started to be used along with FITS, this became a Data Formats BoF, and there was some element of competition between formats, together with an occasional attempt to create a unified Format that could replace FITS as the only astronomical format needed. The impetus for this year's BoF came from an acceptance that astronomy would have to work with multiple formats in the foreseeable future, and a question: Did the use of multiple formats make it difficult to write general-purpose utilities, for example visualisation programs such as SAOImage/DS9, and if so was this a problem? The resulting discussion was interesting, and although it came to no ultimate conclusion, it at least made it clearer why such a conclusion was hard to reach.
\end{abstract}



\section{Introduction, and a little history}

In introductory remarks, the issue was framed as follows:
\vspace{.2cm}

Some people can remember the days before FITS, the wild frontier days when every postdoc had their own data format and their own program to read it. FITS changed all that, and there was a golden age when most astronomical data (and some from the Vatican, which FITS was very proud of) were written in FITS, and, because the format was settled, people could write general-purpose programs like SAOImage that could display almost everyone's images.

But FITS was a `transport' format, and it wasn't easy to manipulate FITS files on disk. So people writing data reduction programs came up with other formats whose data could be modified in place. And some of these implemented a hierarchical format, which provided a lot of flexibility. But they weren't designed for data interchange, and none of them were a standard, so people kept using FITS for that. But FITS couldn't easily handle everything you could put in a hierarchical format, and perhaps that's where the drift from FITS started. And this accelerated with the introduction of formats like HDF. By the time HDF got to HDF5 it had settled down into a de facto standard of sorts. And it's used a lot. In many different ways. But not everyone was comfortable with HDF5, and it's not the only game in town -- ADASS 2024 included an ASDF tutorial, for example.

And the FITS BoFs (e.g. \citet{2019ASPC..523..701M} became data formats BoFs (e.g. \citet{2020ASPC..527..761M}), and people tried -- and failed -- to come up with a new all-encompassing standard. And even if you fix on something really flexible like HDF5, there are so many ways of using it, so many schemas, that just saying this is in such and such a hierarchical format doesn't mean you know where to find the main data array -- if there is one -- and you certainly may not know where to look for its coordinate data.

It looks as if, in the near future, we aren't going to be all agreeing on a universal data standard. So surely that makes it pretty hard to write an SAOImage that can handle any astronomical data file you throw at it.
\vspace{.2cm}

Three questions:
\begin{itemize}
\item Is it the case that it's now difficult to write general purpose programs that can handle any astronomical data?
\item If so, is this a problem? (And let's remember that we may not be a representative sample. Most people here could throw together a purpose-built astropy program to read a given data file and do something with the data in it. But can every astronomer?)
\item If it is a problem, can we do something about it? 
\end{itemize}

As might have been expected, the discussion didn't quite follow that neat and logical sequence.
\section{Discussion}

At first, it seemed there was a consensus that it was now hard to write general-purpose programs able to handle a variety of data files. However, as discussion danced around this topic, an alternative view emerged, led mostly by comments from Erik Tollerud, to the effect that this was not a fundamental problem; all that was really required was to split the problem into layers, and get the data into your system via the appropriate layer.

Although such a description isn't specific to a Python-based system, it was clearly a point of view influenced by, say, an astropy approach, where it is easy to produce a new program by picking and choosing from a large selection of different modules, each with their own capabilities. Certainly, astropy has modules that can read FITS files and modules that can read HDF5 and other formats, and it's possible to mix and match these to produce programs that can take a variety of disparate data formats as input.
\vspace{.2cm}

Does this solve the problem in principle? And does it solve it in practice?
\vspace{.2cm}

Cleary, this sort of approach works best if there are modules that can read all the individual flavours of data file, which begs the question of who provides these modules. And using these modules becomes easier if they all conform to a standard interface that can be used by the higher levels. That may also make them easier to implement, because the writers don't have to decide for themselves what the higher levels would find useful. But that begs the question of how the standard interface gets defined.

The question of defining an interface got a little sidetracked by a question of language. Erik Tollerud pointed out that an interface is easier to specify in a given language, and there was a digression prompted by the claim that in 40 years there would still be a way to run Python. This may well be true -- there are emulators now for some very old games consoles -- and a clear definition remains a clear definition, independent of language, but that may not be the point.

\vspace{.2cm}

A definition of interfaces moved almost inevitably to a discussion of data models; you need a clear picture of what your data contains in order to define an interface to it. This is also connected with the distinction between syntax and semantics -- syntax is a matter of actual layout, that a file can contain arrays, and structures, and values and how these are accessed, while semantics is how they are to be interpreted; which array is the main data array, which keywords define the coordinate transformation, etc.

With people in the room who had been involved in previous attempts to define an overarching model for astronomical data (and the subsequent understandable failure to converge to any final result) it was acknowledged that searching for an all-encompassing model was letting the perfect be the enemy of the good, and discussion moved to whether it would be possible to define a model of a useful subset of data.
\vspace{.2cm}

Mixed in with this was discussion of who would write the access modules that sat between a given data format and the higher levels, the modules that would read that format and make it available using the (still TBD) standard interface. There seemed to be a consensus that these would best be provided by the people responsible for generating the data, and also a feeling that not all those responsible would do so. 

Examples were provided contrasting formats using HDF5 (which essentially defines the syntax) that provided clear schemas explaining the function of the various components (ie the semantics) with others that lacked such clear schemas. However, it was also remarked that locking oneself into a fixed schema, a strict scheme with a tool that enforces it, could get in the way at a later date, and this was perhaps a fundamental problem.

Was it possible to define an abstract data model that described a useful subset of astronomical data? Towards the end of the discussion it was suggested that the community did have a concrete example of a subset data model that was undeniably useful. And this was FITS. Was this a possible, if limited, solution? 

Define an abstract data model that provides access to data arrays, a set of scalar values, and even tables of values, as FITS does, and define an interface -- an API -- to this, implement it for a number of formats, and then you could create a program that could work with any of those formats. It might not make use of everything that format could contain, but you could, for example, display the images, ideally with coordinate information. But then someone burst the bubble by saying; `well, you could just convert the file to FITS and read that in'. Since most people creating non-FITS files do provide at least a minimal FITS conversion program, introducing a complicated software layer that merely did that conversion in situ seems redundant.
\vspace{.2cm}

Since time had also run out, that seemed a natural place to stop the discussion.

\vspace{.2cm}
At least for the moment.

\section{Reflections}
In the end, did this BoF realise that FITS provided a useful minimal concrete data model, but at the same time also demonstrate that creating a program layer and API implementing such a model was unnecessary, because one could always just convert a non-FITS file to FITS anyway?

Arguably, yes. But is there a loss of flexibility here? Perhaps an early question illustrates this. The discussion had considered a program that used a module targeted at a given non-FITS format that implemented a defined interface, which would allow the program to access files in that format. The question was: could this program also read FITS? The simple answer was no. However, the same program could be linked with an equivalent module that implemented the same interface but for FITS files. Having that defined interface meant that, if both modules implementing that interface were available, producing both versions of the program would be trivial, even in a language like C++. In Python it might need no more than another import line.

And that actually sounds useful. It would be even more useful if a single program could be linked with modules for a number of different formats, and yet more useful if the interface also allowed for access to some of those features of hierarchical formats that have drawn people to those formats. 

At a minimum one could imagine that the interface might simplify the view of the data. If it defined a call that provided access to the main data in the file as an N-dimensional array and another that returned the N coordinates for an element of that array, no matter how stored in the format, this would simplify a lot of use cases. Perhaps it could include a call to access the observation details such as exposure time in a standard form such as that defined by ObsCore. \citet{Pxx_adassXXXV} describe the use of mapping files to get such information from FITS files that follow their own rules as to keyword usage, and that concept could be extended to other formats.

It might help to have a standardised way of describing elements of hierarchical files that could be used in such mapping files to indicate the location of such values. This might allow a general module that understood HDF5 syntax in general to use a mapping file to help understand the semantics of a particular schema. (SAOImage has something of this functionality for handling some additional formats, but these do not yet support HDF5.)

The warning about allowing the perfect to be the enemy of the good applies, and it's too easy to get carried away. But maybe a limited interface that was just a little simpler than the way FITS files are accessed and which provided just a little more functionality than FITS might be a useful thing to have? After all, most people in the discussion had seemed to think it sounded like a good idea at first.
\vspace{.2cm}

Perhaps next year.
\section{References}
\bibliography{057}  

\begin{thebibliography}{}
\expandafter\ifx\csname natexlab\endcsname\relax\def\natexlab#1{#1}\fi
\expandafter\ifx\csname url\endcsname\relax
  \def\url#1{\texttt{#1}}\fi
\expandafter\ifx\csname urlprefix\endcsname\relax\def\urlprefix{URL }\fi
\providecommand{\eprint}[2][]{eprint: #2}

\bibitem[{{Barloy} \& {Landais}(2026)}]{Pxx_adassXXXV}
{Barloy}, N., \& {Landais}, G. 2026, in ADASS XXXV, edited by TBD (San
  Francisco: ASP), vol. TBD of ASP Conf. Ser., TBD

\bibitem[{{Mink} et~al.(2020){Mink}, {Diaz}, {Fernique}, {Landais}, {Mireille},
  \& {Michel}}]{2020ASPC..527..761M}
{Mink}, J., {Diaz}, R., {Fernique}, P., {Landais}, G., {Mireille}, L., \&
  {Michel}, L. 2020, in ADASS XXIX, edited by R.~{Pizzo}, E.~R. {Deul}, J.~D.
  {Mol}, J.~{de Plaa}, \& H.~{Verkouter}, vol. 527 of ASP Conf. Ser., 761

\bibitem[{{Mink} et~al.(2019){Mink}, {Diaz}, {Shortridge}, \&
  {Jenness}}]{2019ASPC..523..701M}
{Mink}, J., {Diaz}, R., {Shortridge}, K., \& {Jenness}, T. 2019, in ADASS
  XXVII, edited by P.~J. {Teuben}, M.~W. {Pound}, B.~A. {Thomas}, \& E.~M.
  {Warner}, vol. 523 of ASP Conf. Ser., 701

\end{thebibliography}

\end{document}